# Stimulated Radiative Molecular Association in the Early Solar System. II. Orbital Radii of the Planets and Other Satellites of the Sun

James C. Lombardi Sr. Professor Emeritus,
Physics Department, Allegheny College, Meadville, PA, USA; james.lombardi@allegheny.edu

ABSTRACT

In a previous investigation, the orbital radii of regular satellites of Uranus, Jupiter, Neptune, and Saturn are shown to be directly related to photon energies in the spectra of atomic and molecular hydrogen. To explain these observations a model was developed involving stimulated radiative molecular association (SRMA) reactions among photons and atoms in the protosatellite disks of the planets. In the present investigation, the previously developed model is applied to the planets and important satellites of the Sun. A key component of the model involves resonance associated with SRMA. Through this resonance, thermal energy is extracted from the protosun's protoplanetary disk at specific distances from the protosun wherever there is a match between the local thermal energy of the disk and the energy of photons impinging on the disk. Orbital radii of the planets and satellites are related to photon energies ($E_p$ values) in the spectrum of atomic hydrogen. An expression determined previously is used to relate $E_p$ values to temperatures in the disk. Results indicate the surface temperature of the protosun at the time when the evolution of the planets begins is higher than the surface temperature of a typical T Tauri star. The present investigation offers an explanation for the existence of the asteroid and classical Kuiper belts and predicts that a primordial belt once existed in the vicinity of Neptune. It also indicates that Uranus is formed from two protoplanets and is thus consistent with the theory that the large tilt of Uranus's axis of rotation was created by the collision of two bodies.

## 1. INTRODUCTION

The present investigation is an application of a model (Lombardi 2015) developed to explain interesting relationships among the orbital radii (lengths of the semi-major axis) of regular satellites of the giant gaseous planets in the solar system. In that model the protosatellite disk around each protoplanet has a radially dependent temperature, and regular satellites are ultimately formed where disk temperatures have specific values depending on the energies of the photons impinging upon the disk. In the present application the protosun has a protoplanetary disk that is affected in a similar fashion.

As in Lombardi (2015) TD stands for the *midplane temperature distribution or portion of a temperature distribution in the protoplanetary disk of the Sun where and when the evolution of its planets begins*. The TD for the protoplanetary disk is developed in sections 2.1 and 2.2. Sections 2.3, 2.4, and 2.5 deal with three portions of the entire TD with each section focusing on a belt in the solar system.

The following formula holds for photon energies ($E_p$ values) in the spectrum of atomic hydrogen, with energy in units of cm$^{-1}$, i.e. units with $hc = 1$, where $h$ is Planck's constant and $c$ is the speed of light.

$$E_p(n_f, n_i) = 109737 \text{ cm}^{-1} (1/n_f^2 - 1/n_i^2), \quad (n_f = 1,2,3,\cdots;\ n_i = (n_f+1),(n_f+2),\cdots), \qquad (1)$$

where $n_f$ and $n_i$ are the quantum numbers for the final and initial states associated with the emission of a photon from a hydrogen atom. Photon energies that have the same $n_f$ in Eq. (1), are all in the same series with the series limit $E_p(n_f,\infty)$. In the stimulated radiative molecular association (SRMA) model of Lombardi (2015) the radiation in the spectrum of atomic hydrogen interacts with atoms in the protosatellite disk of Saturn and initiates the evolution of the satellites and rings in Saturn's system. The spectrum of atomic hydrogen plays the same key role in the present investigation with the source of the radiation being the protosun or the protoplanetary disk itself.

In this discussion *resonance ring* refers to a ring shaped region where SRMA resonance exists as introduced in Lombardi (2015). Associated with each resonance ring is a particular $E_p(n_f,n_i)$ value. Radiation emitted from hydrogen atoms initiates resonance in these rings. Matter collects within the rings, as first discussed in Lombardi (2015), and planet evolution begins. Each resonance ring has a change in temperature $\Delta T$ across the ring's width and a radial width $\Delta R$. $\Delta T$ and $\Delta R$ are first evaluated in section 2.3 where the inner and outer edge radii of the asteroid belt are considered.

In Lombardi (2015) SRMA is used to develop the following relationship

$$T_m(n_f,n_i) = 2/3 \ hc/k_B \ (E_p(n_f,n_i) - E_b), \qquad (2)$$

where $T_m(n_f,n_i)$ is the midplane temperature about halfway between the inner and outer radii of a resonance ring, $E_p(n_f,n_i)$ is the energy of a photon associated with the resonance ring, $E_b$ is an adjustable parameter associated with the binding energy of the state in which a molecule is formed during a SRMA reaction, $k_B$ is the Boltzmann constant, and $hc/k_B$ = 1.439 cm kelvin.

## 2. RESULTS and DISCUSSION

*2.1 Matching Orbital Radii with Photon Energies*

Consider Table 1 where orbital radii ($R$ values) of the planets and asteroids are paired with photon energies and midplane temperatures ($E_p$ and $T_m$ values) that are calculated using Eqs. (1) and (2), with $E_b$ equal to 846 cm$^{-1}$. This value of $E_b$ is determined by a method described in section 2.5 as part of the discussion concerning the classical Kuiper belt. As seen below this particular set of pairings produces a TD that is smoothly varying and gives results that match observations. No photon energies have been omitted in the range that the table covers except for those with energies approaching the limits $E_p(9,\infty)$, $E_p(10,\infty)$, and $E_p(11,\infty)$. As described in Lombardi (2015), near a series limit many closely spaced photon energies create overlapping resonance rings and thus a broad circular band of resonance. In Lombardi (2015) such bands evolve into rings of Saturn. In the present investigation bands of resonance evolve into the belts of the solar system. As the radial coordinate decreases within bands, the temperature of the disk increases to a limit, e.g. $T_m(9,\infty)$. Therefore the limit corresponds to the inner edge of a belt.

To establish the particular set of pairings of $R$ and $T_m$ values seen in Table 1, the following considerations are made. First we assume that some series limit corresponds to the inner edge of the asteroid belt and use trial and error to determine the appropriate limit. When we pair the $E_p(9, \infty)$ and its corresponding $T_m(9,\infty)$ with the inner edge of the asteroid belt and systematically pair successive $T_m$ values with available $R$'s, a TD is produced that is smoothly varying. It is also necessary that the process causes the other limiting temperatures, $T_m(10,\infty)$ and $T_m(11,\infty)$, to automatically pair up with inner edges of belts. When $T_m(9,\infty)$ is paired with the inner edge of the asteroid belt, then $T_m(11,\infty)$ is found to correspond to the inner edge of the classical Kuiper belt and $T_m(10,\infty)$ is close to the temperature associated with Neptune's resonance ring. No belt is currently observed near Neptune. However, the observed scattered Kuiper belt objects (SKBO's), whose orbits have large eccentricities and inclinations, appear to have been scattered by Neptune when the solar system was young (Trujillo et al. 2000). The SRMA model therefore predicts the source of the SKBO's. It is a primordial belt situated close to Neptune. Furthermore, Epsilon Eridani, a star that possibly has a system similar to the Sun's is observed to have three belts (Backman et al. 2009). Two of those belts are positioned similarly to the Sun's asteroid and classical Kuiper belts, while the third one is positioned between the other two, similar to the belt predicted by the SRMA model. It may be that common mechanisms act during the early evolution of the solar and the Epsilon Eridani systems to create three belts in each. Most orbital radii listed in Table 1 are experimentally determined. However the inner and outer radii for the belts are determined in sections 2.3, 2.4 and 2.5 below.

*2.2 The TD of the Protoplanetary Disk*

Fig. 1 has three graphs. The graph represented by data points with a solid line smoothly drawn through the points is the TD, a graph of $T_m$ vs $R$ with data taken from Table 1. There are interesting features of this graph:
1. If there were no physical connection between the $E_p$ and $R$ values in Table 1, then the points in Fig. 1 would be randomly distributed and the curve would not be smooth.
2. The inner edge positions of the asteroid and classical Kuiper belts are correctly predicted relative to planet orbital radii and the primordial belt's position is predicted to be in the region of Neptune's orbit.
3. As we systematically associate $E_p$ values with orbital radii, we conclude that two closely spaced $E_p$ values should both be associated with Uranus. This implies that ultimately two protoplanets are formed with close orbital radii. They eventually collide forming the planet Uranus, consistent with the planet's severely tilted axis.
4. There are six large asteroids, Ceres, Vesta, Pallas, Herculina, Eunomia, and Juno, with orbital radii in the range 2.36 – 2.77 AU (The Asteroid Orbital Elements Database 2015). There are no other large asteroids in this region of the asteroid belt. These six are the first, second, third, eighth, ninth and tenth largest in the asteroid belt respectively (Baer 2010, and Baer et al. 2011) and they represent approximately one-half of the total mass in the asteroid belt ( Baer 2010, and Baer et al. 2011, and Pitjeva 2004). Three of the asteroids (Ceres, Pallas and Herculina) have nearly the same size orbital radius; 2.767, 2.771 and 2.770 AU respectively. Assuming these three radii correspond to a single $E_p$ value we take their average to be 2.77 AU. This orbital radius and the three others belonging to Vesta, Eunomia and Juno pair with the four closely spaced $E_p$'s as

        seen in Table 1.
5. In the range of $E_p$ values listed in Table 1, all $E_p$'s are included for $n_f < 9$. Each of these photon energies corresponds to an orbital radius belonging to a planet or large asteroid.
6. If planet migration plays a role in planet evolution then it acts in such a way as to keep the calculated TD smoothly varying.
7. The TD's of Uranus, Jupiter, Neptune and Saturn derived in Lombardi (2015) all have a similar shape to one another. This symmetry does not extend to the protosun's TD.

The other two graphs in Fig. 1 are midplane temperature distributions for the accretion disk (protoplanetary disk) of a T Tauri star calculated by D'Alessio, Canto, Calvet, Lizano, (1998), henceforth referred to as DCCL. In the DCCL model, for disk radii > 2 AU, the heating of the disk is determined mainly by the irradiation of the star. Inside 2 AU viscous heating is dominant. The curve indicated with short dashes gives the DCCL midplane temperature distribution for a nonirradiated disk, i.e. one whose temperature is derived solely from viscous heating. The longer dashed line gives the DCCL midplane temperature distribution for a disk that experiences viscous heating and is also irradiated by a central star with a surface temperature of 4000 K, a typical temperature for a T Tauri star.

An interesting relationship among the three distributions in Fig. 1 is revealed when the "nonirradiated" DCCL distribution is subtracted from the other DCCL distribution and also from the TD. The two resulting distributions are given in Fig. 2. They compare the effect of the protosun's irradiation of its disk in the present investigation (top graph) with the effect of the protosun's irradiation of its disk in the DCCL investigation (bottom graph). Because the TD graph in Fig. 1 is a series of points, the top graph in Fig. 2 is also a series of points, with the dashed line that connects these points added to guide the eye. The curves in Fig. 2 have similar characteristics. In both cases the graphs rise to their maximum values at about 2-3 AU. From 2 AU out to 25 AU both graphs fall off in a similar way and the temperatures in the top graph are about 13.5 times higher than corresponding temperatures in the bottom graph. This indicates that at the time when the evolution of the planets begins, the protosun's surface is considerably hotter than the surface of a typical T Tauri star. There are subclasses of T Tauri stars called FUors and EXors (Hartmann et al. 1993) that are characterized by short-lived magnitude changes of up to 6 magnitudes with rise times on the order of months to years (Appenzeller and Mundt 1989, and Herbig 1977). These results indicate that the evolution of the planets begins during an amplitude outburst. In Fig. 2 the top graph falls off more rapidly beyond 25 AU. This falloff may be related to the findings of DCCL that their model disk becomes optically thin beyond 20 AU. The optically thin region would not be heated as readily by an outburst of the the central star.

The similarity of the two graphs in Fig. 2 indicates it may be possible to use the theory presented in the DCCL paper and the TD in Fig. 1 to determine the surface temperature of the protosun at the time when the evolution of the planets begins.

*2.3 The Inner TD*

The $T_m$ and $R$ values in Table 1 from Mercury to Jupiter are used to make Fig. 3, the inner TD. In this figure the filled circles are the points for the planets and the four open circles are the points for the six asteroids mentioned above. The radii for the inner and outer edges of the asteroid belt are calculated below and the crosses in Fig. 3 are the points for these edges. The curve in the figure is the best fit through the planet points excluding Mercury's point. Mercury is not fitted well by the line possibly because of its proximity to the inner edge of the protoplanetary disk. The equation of the curve is

$$T_m = 497.1/R^{0.1676} + 45.03 \text{ (Kelvin)}, \qquad (3)$$

where $R$ is in AU. The fitted points fall almost exactly on the curve. The asteroid points are close to the curve with Vesta's being squarely on it. Eq. (3) is used in the determination of the inner and outer edge radii of the asteroid belt.

In the SRMA model the asteroid belt exists because of the infinitely many different photon energies in the series of photons that have the limit $E_P(9,\infty)$. This limit corresponds to the inner edge radius of the belt. Photons with energy values near the limit $E_P(9,\infty)$ create many overlapping resonance rings which combine to form a continuous region of resonance within the boundaries of the belt. Here thermal energy is extracted, matter collects, and ultimately asteroids are formed. Within the asteroid belt, the distance between adjacent resonance rings increases as the radial coordinate increases and eventually the resonance rings do not overlap. The radial coordinate where overlap ceases defines the outer edge of the asteroid belt. But why are there no satellites orbiting just beyond the asteroid belt which are associated with photon energies in the $n_f = 9$ series? A possible explanation involves the emission intensities in atomic hydrogen which are generally lower for higher values of $n_f$ (Wiese and Fuhr 2009). Possibly the intensities for which $n_f < 9$ are strong enough to create isolated resonances rings. But for $n_f \geq 9$, intensities are not strong enough to cause resonance unless resonance rings spatially overlap and mutually support each other.

The outermost feature of the asteroid belt is a distribution of asteroids called the Hilda group. The distribution extends to about 4.15 AU (The Asteroid Orbital Elements Database 2015) which we assume defines the observed outer edge radius of the asteroid belt. To fit this observation we use various $E_P(9, n_i)$ values to calculate trial values for the outer edge radius $R_{OE}$ of the belt. For $E_P(9,46) = 1302.9$ cm$^{-1}$, we find $R_{OE} = 4.11$ AU in good agreement with the observed value. The following paragraph discusses the calculation of $R_{OE}$.

Fig. 4 is a diagram illustrating three portions of adjacent resonance rings with the radial direction to the right. The ring in the middle overlaps the ring on the left. Resonance does not develop in the ring on the right because it does not overlap sufficiently with the middle ring. So the right edge of the middle ring is the outer edge of the asteroid belt and the ring on the left is the last interior resonance ring. To determine $R_{OE}$ we use Fig. 4 and the outline below.

1. The three dashed lines in Fig. 4 are near the middle of each resonance ring. The midplane temperatures at these positions are $T_m(9,47)$, $T_m(9,46)$ and $T_m(9,45)$, determined using Eqs. (1) and (2).
2. $\Delta T$ is the change in temperature across the resonance ring at the outer edge of the asteroid belt (the ring in the middle of the diagram). From the diagram
$\Delta T = [T_m(9,45) - T_m(9,47)]/2 = -2.2$ K.
(The negative indicates the temperature decreases as the radial direction increases.)
3. The temperature at the outer edge of the asteroid belt is
$T_{OE} = T_m(9,46) + \Delta T/2 = T_m(9,46) - 1.1$ K $= 437.3$ K
and from Eq. (3) $R_{OE} = 4.11$ AU
4. $\Delta R$ is the radial width of the resonance ring that is at the outer edge of the asteroid belt.

To determine $\Delta R$, we first determine the temperature of the ring's inside edge is $T_m(9,46) - \Delta T/2$. The ring's inside edge radius is then found from Eq. (3) and is subtracted from $R_{OE}$ to yield $\Delta R = 0.13$ AU.

To determine the inner edge radius of the asteroid belt $R_{IE}$, first we determine the inner edge temperature $T_{IE} = T_m(9,\infty) - \Delta T/2$, where $T_m(9,\infty)$ is found from Eq. (2) and $\Delta T$ is assumed to be – 2.2 K as it is on the outer edge of the asteroid belt. Then $T_{IE}$ in Eq. (3) gives $R_{IE} = 1.96$ AU. This value for $R_{IE}$ is in good agreement with the observed value near 2.0 AU (The Asteroid Orbital Elements Database 2015).

Table 2 contains the key results found in the above calculations plus similar results determined later for the other belts in the protosun's TD. In the calculations for the two other belts, we assume that $n_i$ is 46 for each belt's outer edge, just as it is for the asteroid belt. As we see later this is equivalent to assuming $\Delta T$ has the same value (– 2.2 K) at the outer edge of the other belts too. Because the calculated inner and outer edge radii of the asteroid belt and classical Kuiper belt are close to the observed values, $\Delta T$ most likely is constant or close to constant throughout the TD. On the other hand, because $\Delta T/\Delta R$ is a measure of the local temperature gradient, $\Delta R$ has different values throughout the TD.

*2.4 The Middle TD*

The $T_m$ and $R$ values in Table 1 from Jupiter to Pluto are used to plot the middle TD in Fig. 5. The best fit curve to the TD is given by

$$T_m = 497.1/R^{0.1676} + 45.03 - 0.1356 R^2 \text{ (Kelvin)}. \quad (4)$$

Eq. (4) differs from Eq. (3) by the added last term. When fitting the data, the coefficient of the last term is the only parameter varied.

The inner and outer edge points for the primordial belt that the SRMA model predicts to have existed near Neptune are indicated on the curve in Fig. 5. The method used to determine the coordinates of these points is the same as the one used for the asteroid belt except Eq. (4) is used instead of Eq. (3). The inner edge is associated with the limit $E_P(10,\infty) = 1097.37$ cm$^{-1}$. We assume that the outer edge corresponds to the photon energy $E_P(10,46) = 1045.5$ cm$^{-1}$ and find that again $\Delta T = -2.2$ K. The radial width $\Delta R$ of the resonance ring on the outer edge of the primordial belt is 0.22 AU. The values for the inner and outer edge radii for the primordial belt are found to be $R_{IE} = 25.9$ AU and $R_{OE} = 31.4$ AU respectively. All of these results are listed in Table 2.

*2.5 The Outer TD*

The $T_m$ and $R$ values in Table 1 from Pluto to the outer edge of the classical Kuiper belt are used to plot the outer TD in Fig. 6. The filled circles in this figure are the $(T_m, R)$ points for the dwarf planets Pluto, Haumea and Makemake. Assuming $T_m$ varies exponentially within the classical Kuiper belt, the best fit curve to the three dwarf planet points is given by

$$T_m = 112.8\exp(-0.2885(R-39)) + 4.0 \quad \text{(Kelvin)}. \qquad (5)$$

The inner and outer edge points for the classical Kuiper belt are indicated on the curve in Fig. 6. The method used to determine the inner and outer radii of the classical Kuiper belt is the same as the one used for the inner and middle TD's except in this case Eq. (5) is used. The inner edge is associated with the limit $E_P(11,\infty) = 906.9$ cm$^{-1}$. We assume the outer edge of the classical Kuiper belt corresponds to the photon energy $E_P(11,46) = 855.1$ cm$^{-1}$ and once again find $\Delta T = -2.2$ K. The radial width $\Delta R$ of the resonance ring on the outside edge of the classical Kuiper belt is 1.59 AU. The values for the inner and outer edge radii for the classical Kuiper belt are $R_{IE} = 41.5$ AU and $R_{OE} = 50.9$ respectively. All of these results are listed in Table 2.

The value of $R_{IE}$ is in good agreement with the observed value of 41 AU (Trujillo et al. 2001), and $R_{OE}$ is in good agreement with the observed values of 47 AU (Trujillo et al. 2001) and 50 AU (Jewitt et al. 1998). Jewitt et al. (1998) carried out one of the first surveys to discover Kuiper belt objects. They report "We are uncomfortable with the notion that the Kuiper belt might have an edge near 50 AU (what physical process could be responsible?) but our data nevertheless suggest this as a possibility." The SRMA model provides an explanation for the edge of the classical Kuiper belt near 50 AU.

The value of $E_b$ in Eq. (2) is 846 cm$^{-1}$ for all calculations in this investigation. This value is determined while fitting the outer TD in the following way. We would like the exponential in Eq. (5) to be asymptotic to the cosmic microwave background radiation temperature (CMBRT). Assuming the CMBRT varies inversely with time to the 2/3 power (Carroll and Ostlie 2007) and also assuming reasonable values for the age of the universe and age of the solar system, we calculate this temperature to be about 4 K when the solar system was born. Notice that the constant on the right side of Eq. (5) is 4 K and therefore $T_m$ approaches 4 K for increasing $R$. For instance when $E_b$ is taken to be 848 cm$^{-1}$ rather than 846 cm$^{-1}$, the $T_m$ values calculated from Eq. (2) are about 2 K lower than the $T_m$'s given in Table 1. This does not change the shape of the fits in Figs. 3, 5 and 6 nor does it affect the calculated radii of the inner and outer edges of the three belts. However with $E_b$ equal to 848 cm$^{-1}$, the fit to the outer TD (Eq. (5)) is asymptotic to approximately 2 K rather than 4 K. Therefore we take $E_b$ to be 846 cm$^{-1}$.

*2.6 Makemake and Other Dwarf Planets*

At first it seems there is no photon in the hydrogen spectrum that corresponds to the dwarf planet Makemake. Interestingly, the photons that generate Makemake's resonance ring belong to the same series as those resulting in the asteroid belt (i.e. $n_f = 9$). As proposed earlier for the asteroid belt, resonance does not occur for $n_i <$ approximately 46, because hydrogen transition intensities are not strong enough to cause resonance unless there is spatial overlap among the resonance rings. The radius beyond which there is no more overlap defines the outer edge of the asteroid belt. But there are three more photon energies in the $n_f = 9$ series that create resonance overlap in the other two belts. Two of these create resonance in the primordial belt and one, corresponding to photon energy $E_P(9,15) = 867.1$ cm$^{-1}$, creates Makemake's resonance in the classical Kuiper belt. As mentioned earlier, the emission intensities in atomic hydrogen are generally lower for higher values of $n_f$ (Wiese and Fuhr 2009). For the photon energy of Makemake's resonance ring $n_f = 9$, but for the other resonance rings in the classical Kuiper belt $n_f = 11$. Makemake's resonance is therefore stronger than other resonances in the classical Kuiper belt and it results in the formation of the dwarf planet.

The two $n_f = 9$ photon energies that cause resonance overlap in the primordial belt are $E_P(9,20) = 1080.4$ cm$^{-1}$ and $E_P(9,19) = 1050.8$ cm$^{-1}$. From Eq. (2) these correspond to $T_m = 224.9$ K and $T_m = 196.5$ K. These temperatures in Eq. (4) give $R = 27.8$ AU and $R = 30.8$ AU. This places the "predicted" objects within the boundaries of the primordial belt. As mentioned before, the belt near Neptune does not exist anymore owing to scattering of its members by Neptune. One of the predicted objects might be Eris. Eris is a scattered Kuiper belt object and the largest of the five dwarf planets. If one of these objects is in fact Eris, then the SRMA model accounts for all five of the objects presently classified as dwarf planets, i.e. Ceres, Pluto, Haumea, Makemake, and Eris.

## 3. FURTHER DISCUSSION

As discussed in Lombardi (2015) the temperature and pressure would tend to decrease within a resonance ring once resonance begins. Then matter moves into the ring and a gravitational instability is created and ultimately a planet is formed. Indeed, disk instabilities (but without the SRMA trigger) have been proposed (Boss 2000 & Helled et al. 2014) partly to explain the discrepancy between protoplanetary disk lifetimes and the theoretical longer times required for planets to develop within the disk.

The author thanks James Lombardi Jr. and H. Gerald Staub for helpful suggestions on the manuscript for this paper, and Allegheny College for use of computer facilities.

**Table 1.** Orbital radii, photon energies and midplane temperatures associated with planets, other satellites and belts in the solar system.

|  | R(AU)[a,b] | ($n_f$, $n_i$) | $E_p(n_f, n_i)$[c] | $T_m(K)$[d] |
|---|---|---|---|---|
| Mercury | 0.387 | (8,21) | 1465.8 | 594.6 |
| Venus | 0.723 | (8,20) | 1440.3 | 570.1 |
| Earth | 1.000 | (8,19) | 1410.7 | 541.7 |
| Mars | 1.524 | (8,18) | 1375.9 | 508.4 |
| IE[e] Asteroid Belt | 1.960 | (9,∞) | 1354.8 | 489.2 |
| Vesta | 2.362 | (5,6) | 1341.2 | 475.1 |
| Eunomia | 2.644 | (8,17) | 1334.9 | 469.0 |
| Juno | 2.671 | (6,8) | 1333.6 | 467.8 |
| Ceres, Pallas & Herculina | 2.770 | (7,11) | 1332.6 | 466.8 |
| OE[e] Asteroid Belt | 4.110 | (9,46) | 1302.9 | 437.3 |
| Jupiter | 5.203 | (8,16) | 1286.0 | 422.1 |
| Saturn | 9.537 | (8,15) | 1226.9 | 365.4 |
| Uranus | 19.190 | (8,14) | 1154.8 | 296.2 |
| Uranus | 19.190 | (7,10) | 1142.2 | 284.2 |
| IE[e] Primordial Belt | 25.900 | (10,∞) | 1097.4 | 242.2 |
| Neptune | 30.070 | (8,13) | 1065.3 | 210.4 |
| OE[e] Primordial Belt | 31.400 | (10,46) | 1045.5 | 190.3 |
| Pluto[f] & Plutinos | 39.480 | (8,12) | 952.6 | 102.3 |
| IE[e] Classical Kuiper Belt | 41.500 | (11,∞) | 906.9 | 59.5 |
| Haumea | 43.243 | (7,9) | 884.8 | 37.2 |
| Makemake | 45.729 | (9,15) | 867.1 | 20.2 |
| OE[e] Classical Kuiper Belt | 50.900 | (11,46) | 855.1 | 7.61 |

[a] The Asteroid Orbital Elements Database (2015)
[b] National Space Science Data Center (2015)
[c] Calculated with Eq. (1)
[d] Calculated with Eq. (2)
[e] IE and OE stand for inner edge and outer edge. IE and OE radii of belts are determined in sections 2.3, 2.4 and 2.5.
[f] Pluto's orbital radius is used in the analysis.

**Table 2.** Calculated inner and outer radii of belts in the protosun's TD. Also listed are the radial width and temperature change across the width of the resonance ring near the outer edge of each belt.

|  | $R_{IE}$ | $R_{OE}$ | $\Delta T(K)$ | $\Delta R(AU)$ |
|---|---|---|---|---|
| asteroid belt | 1.96 | 4.11 | -2.2 | 0.13 |
| primordial belt | 25.9 | 31.4 | -2.2 | 0.22 |
| classical Kuiper belt | 41.5 | 50.9 | -2.2 | 1.59 |

Figure 1. The protosun's TD compared to temperature distributions calculated for an irradiated and a non-irradiated disk (D'Alessio, Canto, Calvet, Lizano (DCCL) 1998). The TD: filled circles and solid line; the irradiated disk: long dashed line; the non-irradiated disk: short dashed line. Also indicated on the TD are planets, dwarf planets and large asteroids in the solar system. The open circles, squares, and diamonds indicate the inner and outer edges of the three belts.

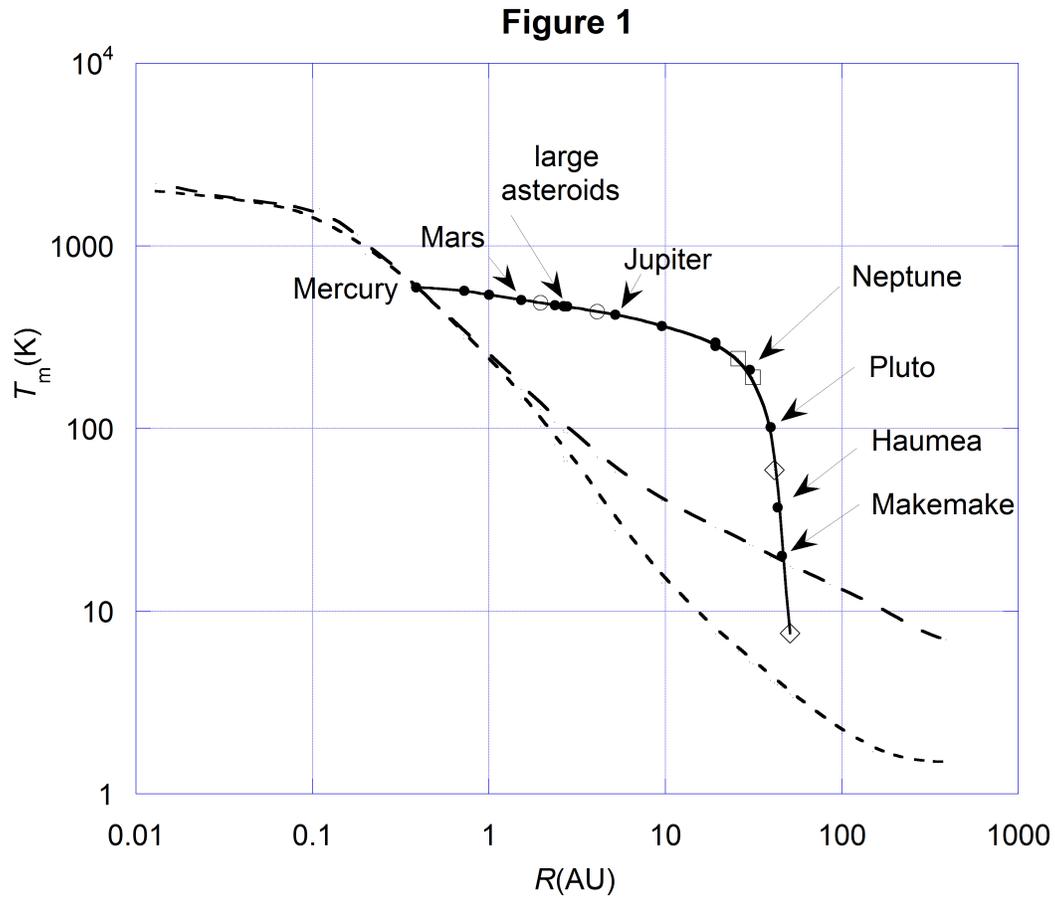

Figure 2. Difference distributions derived from the protosun's TD and the DCCL temperature distributions for irradiated and non-irradiated disks. TD minus "non-irradiated": filled circles and dashed line; "irradiated" minus "nonirradiated": solid line. Also indicated on the TD are planets, dwarf planets and large asteroids in the solar system.

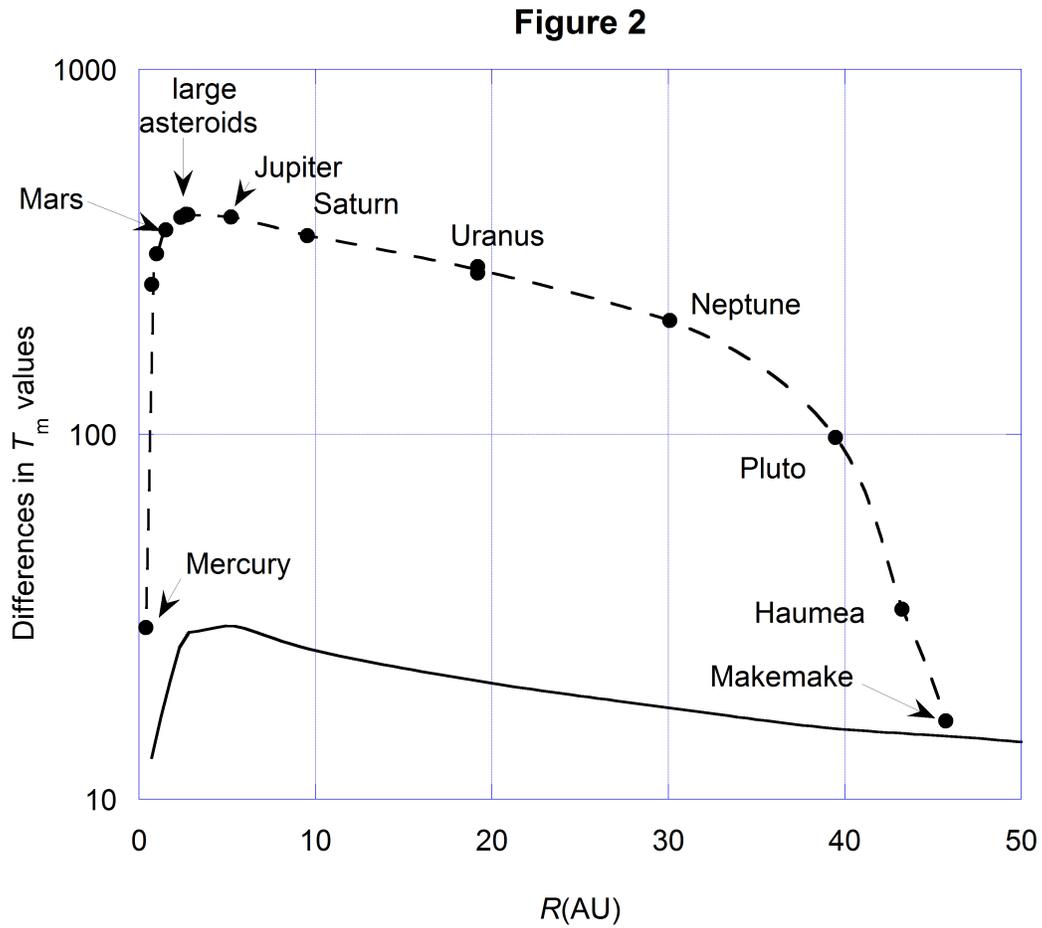

Figure 3. The inner TD. Planets (filled circles) and large asteroids (open circles) are indicated. The calculated inner and outer edge radii of the asteroid belt are $R_{IE}$ = 1.96 AU and $R_{OE}$ = 4.11 AU.

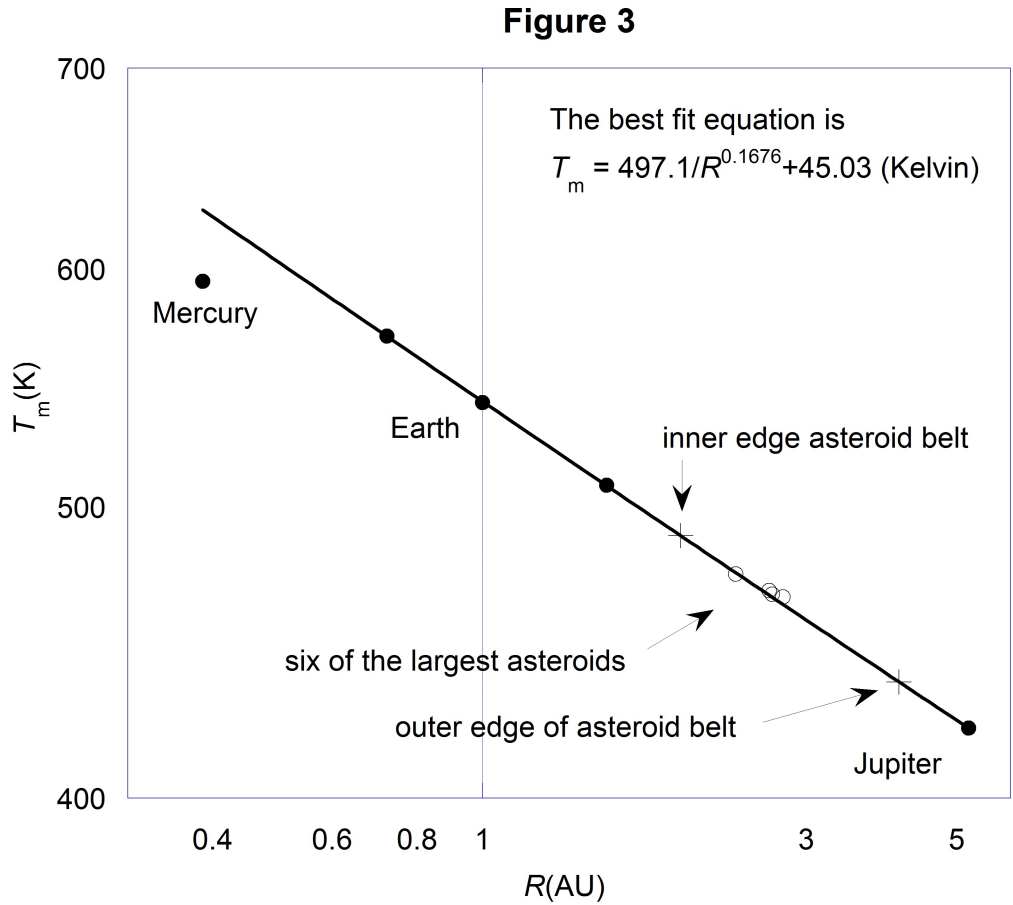

Figure 4. The diagram used to calculate the outer edge radius $R_{OE}$ of the asteroid belt, $\Delta T$ and $\Delta R$.

**Figure 4**

Determining the Outer Edge Radius ($R_{OE}$) of the Asteroid Belt

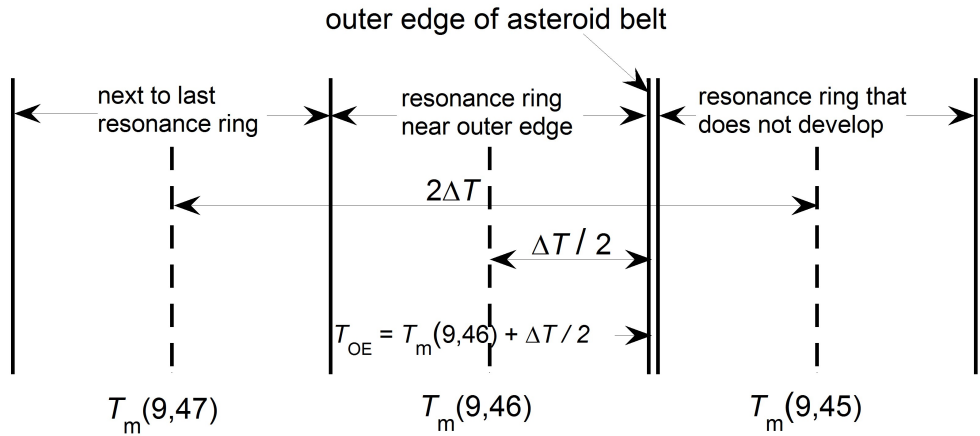

$T_{OE}$ in Eq.(3) gives $R_{OE}$ = 4.11 AU

Figure 5. The middle TD. The filled circles indicate the points for the giant gas planets and the dwarf planet Pluto. The two points for Uranus indicate that this planet was formed by the collision of two protoplanets. The calculated inner and outer edge radii of the primordial belt are $R_{IE} = 25.9$ AU and $R_{OE} = 31.4$ AU.

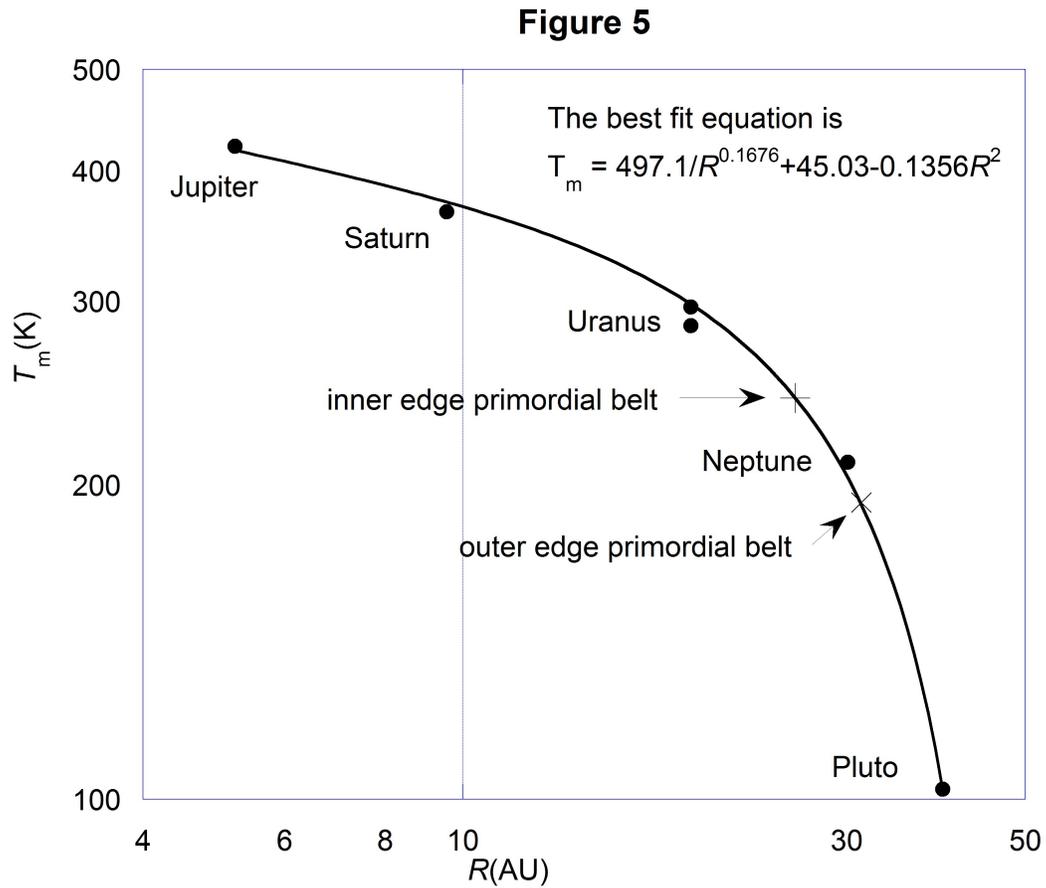

Figure 6. The outer TD. The filled circles indicate the points for the the dwarf planets Pluto, Haumea and Makemake. The calculated inner and outer edge radii of the classical Kuiper belt are $R_{IE}$ = 41.5 AU and $R_{OE}$ = 50.9 AU.

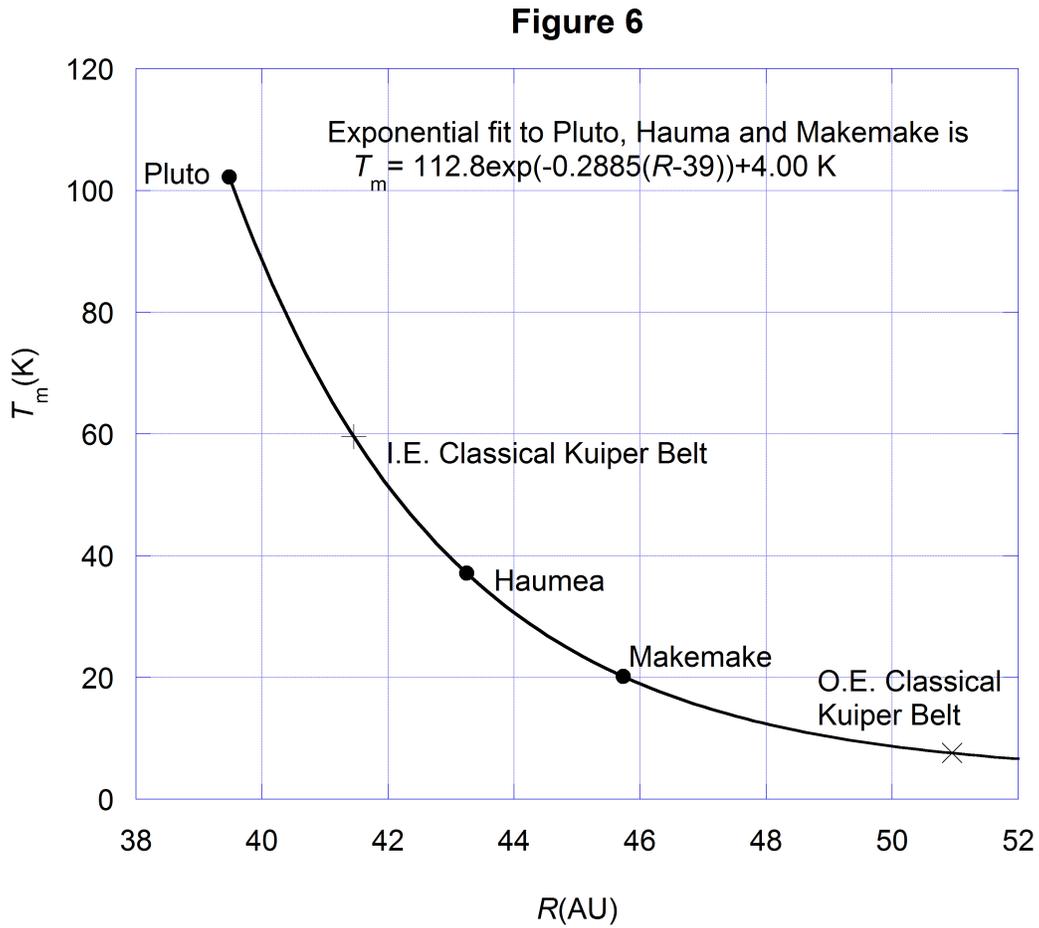